\documentclass[aps,prb,twocolumn,showpacs,superscriptaddress,floatfix]{revtex4-2}

\usepackage{graphicx,graphics}
\usepackage{dcolumn}
\usepackage{amsmath,amssymb,amsfonts}
\usepackage{latexsym,verbatim}
\usepackage{bm}
\usepackage{color}
\usepackage{ulem}
\usepackage[breaklinks=true,colorlinks,citecolor=blue,linkcolor=blue,urlcolor=blue]{hyperref}
\usepackage[percent]{overpic}
\usepackage{bbold}
\usepackage{amsmath}
\usepackage{footnote}
\usepackage{comment}
\usepackage{epsfig}
\usepackage{float} 
\graphicspath{{./images/}}
\begin{document}
\title{Impact of structural defects on the performance of graphene plasmon-based molecular sensors}
\author{Karina A. Guerrero-Becerra}
\affiliation{Istituto Italiano di Tecnologia, Via Morego 30, I-16163 Genova,~Italy}
\author{Remo Proietti Zaccaria}
\affiliation{Istituto Italiano di Tecnologia, Via Morego 30, I-16163 Genova,~Italy}
\begin{abstract}

Graphene-based plasmonic devices are regarded to be suitable for a plethora of applications, ranging from mid-infrared to terahertz frequencies. In this regard, among the peculiarities associated with graphene, it is well known that plasmons are tunable and tend to show stronger confinement as well as a longer lifetime than in the noble-metal counterpart. However, due to the two-dimensional specificity of graphene, the presence of defects might induce stronger effects than in bulky noble metals. Here, we theoretically investigate the impact of structural defects hosted by graphene on selected figures of merit associated to localized plasmons, which are of key technological importance for plasmon-based molecular sensing. By considering an optimized graphene nanostructure, we provide a comparative analysis intended to shed light on the impact of the type of defect on graphene localized plasmons, that regards distinct types of defects commonly arising from fabrication procedures or exposure to radiation. This understanding will help industry and academia in better identifying the most suitable applications for graphene-based molecular sensing.
\end{abstract}
\maketitle
\section{Introduction}
Collective electronic excitations hosted by graphene and by other 2D materials, as well as van der Waals heterostructures, have been found to advance protocols for molecular sensing based on the interaction with analytes~\cite{Hu_2019}. The advancement relies on the fact that 2D materials used as plasmonic substrate for molecular sensors display large adsorption capacity due to their maximal surface-to-volume ratio and host collective excitations with proven figures of merit (f.o.m.)~\cite{Low_2017}.
In particular, graphene-based structures, employed as plasmonic substrate materials, have been shown to increase the sensitivity of surface-enhanced infrared absorption (SEIRA) spectroscopy in the label-free detection of analytes such as proteins~\cite{Rodrigo_2015,Wu_2017}, and biomolecules in aqueous phase~\cite{Zheng_2018}. Furthermore, plasmonic graphene-based structures enable quantitative bioassay~\cite{Bareza_2022}, multi-resonant SEIRA spectroscopy~\cite{Cai_2019}, and highly selective gas sensors~\cite{Bareza_2020}, thus extending the functionalities of traditional SEIRA spectroscopy~\cite{Hu_2019}. 
Finally, novel molecular sensing protocols based on plasmonic enhancement, rely on graphene used as plasmonic substrate to overcome the weak absorption of gas species~\cite{Khaliji_2019} and the encumbrance of spectrometers or laser sources, hence promoting device miniaturization~\cite{Marini_2015, Guerrero_2019}.
In order to capitalize on the advantages that 2D materials can bring to the molecular sensors sector, it is important to consider the effect of unavoidable structural defects on the f.o.m of plasmons. 
Especially for graphene, intrinsic structural defects~\cite{Tian_2017} are native or are physically introduced after fabrication procedures, device processing~\cite{Banhart_2011} or exposure to environmental factors, such as radiation exposure~\cite{Krasheninnikov_2020}.
Defects need to be considered as they change the optical, electronic, mechanical, thermal and chemical properties of graphene-based structures~\cite{Tian_2017,Vicarelli_2015,Bhatt_2022}.
Changes in the properties of pristine graphene caused by defects might non-trivially depend on the defect type. In this respect, while different types of structural defects diversely impact the electronic properties of bulk graphene, as recently shown in Refs.~\onlinecite{Kot_2020},
a comparative analysis of their impact on graphene plasmons has not been realized yet.
Here, we analyze the impact on selected f.o.m as concerns graphene localized plasmons of three types of intrinsic atomic-size structural defects that are both natural and induced by radiation exposure~\cite{Bhatt_2022}: Stone-Wales defects (\textit{SW}), single vacancies (\textit{sv}) and double vacancies (\textit{dv}). Two f.o.m of graphene localized plasmonic resonances, being critical requirements for graphene to succeed in molecular sensing applications, are on the focus here: the quality factor and the dynamic tunability. 
The quality factor impacts the sensitivity of molecular sensors that rely on plasmonic-enhancement: a higher quality factor promotes a longer light–matter interaction time between the plasmons and the analyte, hence inducing higher enhancement of the electric field~\cite{BinAlam_2021,Tanaka_2021}. Therefore, it is an important figure of merit for designing sensors based on SEIRA spectroscopy and other sensing protocols based on plasmonic enhancement. 
The dynamic tunability refers instead to the possibility to tune the resonant energy of graphene plasmonic excitations, and derivatives, by tuning a gate voltage or by employing chemical doping, hence enlarging the range of potential applications. Within the context of plasmon-based sensing protocols, the dynamic tunability allows for setting the plasmonic excitations to overlap with different molecular bands. This is an important advantage with respect to traditional (metallic) substrates for plasmon-based molecular sensing, for which the same functionality is limited~\cite{Oh_2021}. Indeed, the lack of post-fabrication tunability of plasmons in metals, and their limited range of energies covered by the spectral width of each individual plasmon, imply that metallic plasmons are able to enhance only a few of the analyte excitations. In order to identify molecular excitations covering different spectral portions, it is necessary to change the metallic plasmonic substrate. In contrast, after fabrication, graphene plasmon energy can be tuned electrically over the entire molecular fingerprint region~\cite{Hu_2016}. Furthermore, graphene plasmon energy can be tuned through reversible chemical doping, in a scheme applicable as molecular sensing protocol itself~\cite{Bareza_2020}.
Chemical tuning bears the advantages linked to a free-gate structure, e.g. a transparent structure.  Specifically, a strongly localized plasmon can be sustained at zero bias, while a gated structure requires a high bias for the same plasmonic response~\cite{Paulillo_2021}. The possibility to tune the optical response in graphene with electrical or chemical doping enables frequency-selective enhancement of molecular bands~\cite{Rodrigo_2015}, extends the typical narrow range of frequency of individual plasmons in metals, and opens the path towards the achievement of multi-band detection through a single device~\cite{Hu_2016}. 
In this respect, by considering  an equilateral triangle shaped graphene nanostructure, here we investigate the impact of some of the prevalent structural atomic-scale defects on the quality factor and on the dynamic tunability. 
The choice of the triangular geometry is driven by its well defined localized plasmons, its large pristine quality factor, and the proven strong resilience to single vacancies~\cite{Aguillon_2021}. Furthermore, Ref.~\onlinecite{Aguillon_2020} shows that a nanoantenna based on the triangular geometry realizes strongly localized hotspots providing benefits for molecular detection.
\section{Methodology}
The impact of atomic-scale defects on localized plasmons hosted by a graphene nanostructure is studied through our numerical implementation of the Quantum Tight-Binding Random Phase Approximation (Q-TB+RPA) method~\cite{Thongrattanasiri_2012,Cox_2016,Wang_2015}. This is a fully nonlocal and atomistic quantum mechanical treatment of the collective plasmonic excitations that goes beyond the widely used harmonic approximation. In particular, the Q-TB+RPA method enables the detailed description of the impact on the plasmonic spectrum of nanoscale-size details of the atomic structure such as edge conformation~\cite{Wang_2015} and defects~\cite{Aguillon_2021}.
Within this method, the electronic structure is described by a tight-binding (TB) model that carries the information on the atomic structure. The TB model here considered is a minimal nearest neighbor model that takes the approximation of ignoring the relaxation of the atomic positions after defect formation, which has been treated elsewhere~\cite{Tian_2017}.
Within the TB model, the hopping energy after defect formation is described through a strain model. The electronic structure is an input for the calculation of the response function. The dielectric function is computed at the level of the random phase approximation (RPA)~\cite{Giuliani_2005}, while full non locality of the response is considered, as dictated by the small size and by the lack of translational symmetry of the nanostructure. The plasmonic spectra over a selected range of energy  $\hbar \omega$ is described in terms of the loss function  $-\Im[1/\varepsilon_{n_{1}}(\omega)]$, obtained from the eigenvalues, $\varepsilon_{n_{1}}(\omega)$, of the dielectric function. The localized plasmons, defined as the peaks of the loss function corresponding to zeroes of the dielectric function, are numerically calculated. The theoretical model is fully described in Appendix (\ref{app:formalism}).
In the results shown below, the plasmon spectra have been obtained by setting the temperature at $T = 300 \, {\rm K}$ and the chemical potential at  $\mu = 1.6 \, {\rm eV}$ (unless otherwise stated). These parameters determine the occupation number of electronic states as described by the Fermi-Dirac distribution function, Eq.~(\ref{eq:fddistri}), and through it, the density-density response function,  Eqs.~(\ref{eq:cgmatrix}) and~(\ref{eq:chi}). The latter response function depends also on the inverse relaxation time, here being fixed at $\eta = 6 \, {\rm meV} / \hbar$~\cite{Thongrattanasiri_2012,Westerhout_2018}. The plasmonic spectra depend on the dielectric environment through the Coulomb interaction described in Eq.~(\ref{eq:coulomb}), where the relative permittivity is set at $\epsilon_{r} = 1$. With these values of parameters, the dielectric function in Eq.~(\ref{eq:epsilon}) is fully determined, therefore the plasmonic spectra.
\section{Numerical results}
\begin{figure*} 
\begin{overpic}[width=0.98\linewidth]{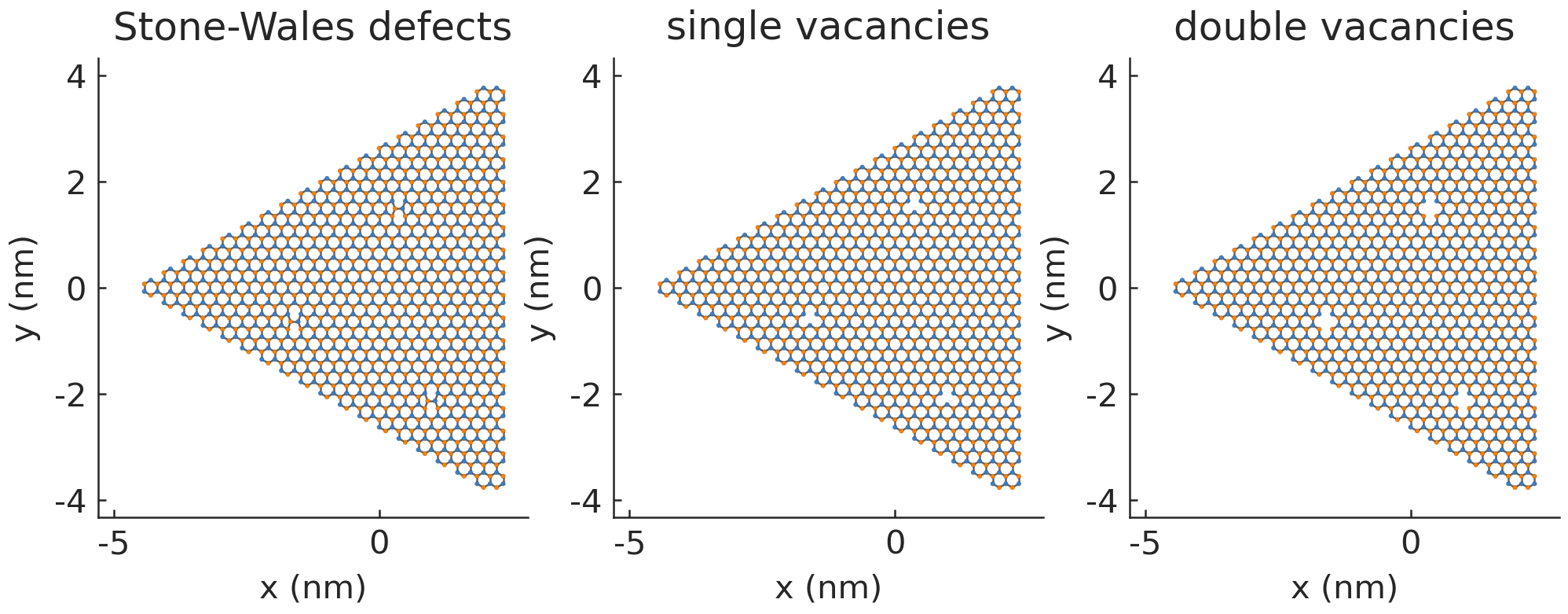}\put(0.7,39.0){(a)}\end{overpic}
\begin{overpic}[width=0.49\linewidth]{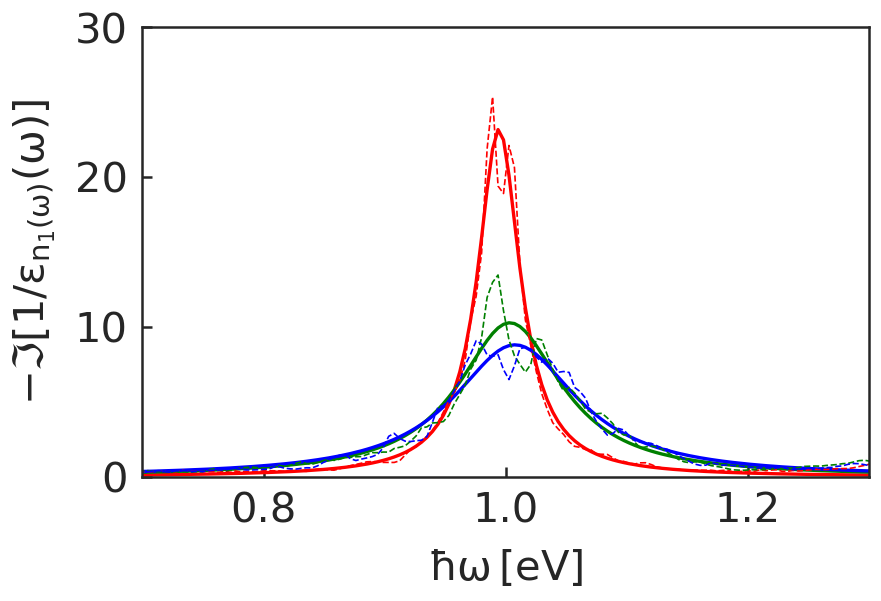}\put(2,65){(b)}\end{overpic}
\begin{overpic}[width=0.49\linewidth]{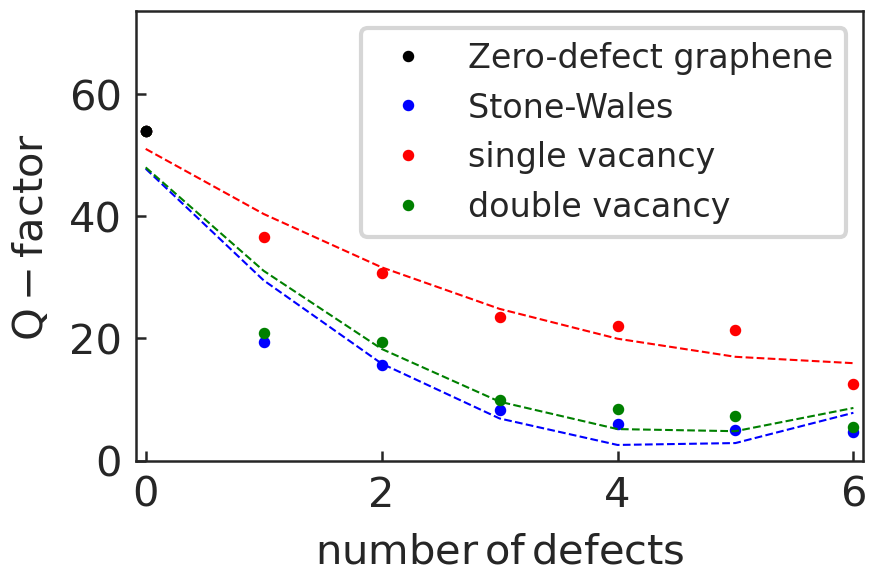}\put(2,65){(c)}\end{overpic}
\caption{\label{fig:qfactormain}
Panel (a): atomic structures of a graphene nanostructure hosting three types of structural defects. Panel (b): comparison of the loss function as a function of the energy for different types of structural defects. The plasmonic peak (dashed line) and Lorentzian fit (solid line) corresponding to the (three) defects in panel a) are displayed with color code: Stone-Wales (\textit{SW}, blue), single vacancies (\textit{sv}, red), and double vacancies (\textit{dv}, green). Panel (c): quality factor as a function of the number of defects. The dashed lines (a guide to the eye) represent the quadratic interpolation of the data points. The same color code as in panel b) distinguishes the type of defect. The values reported for 3 defects correspond to the configurations in panel a).}
\end{figure*}
We consider a pristine triangular graphene nanostructure formed by carbon atoms with minimum two bonds and characterized by armchair edges, the overall structure is constituted by $N = 1098$ atoms, corresponding to a triangle side of $7.5$ nm. The pristine graphene nanostructure was then decorated with different kinds of defects, in particular Stone-Wales, single vacancies and double vacancies defects, as shown in Fig.\ref{fig:qfactormain}(a).
With the aim of comparing the impact of these kinds of structural defects on the optical response of the graphene nanostructure, we select a peak of the plasmonic spectrum of the pristine graphene nanostructure and monitor its change upon insertion of the defects. 
In order to consistently describe the peak width, we apply a fitting procedure based on a single Lorentzian function and extract from it the full width at half maximum (FWHM). This approach was employed for hindering the effects of the small kinks that can be seen in Fig.\ref{fig:qfactormain}(b), possibly due to the limited dimensions of the graphene structure here considered which determines the splitting of a broad peak into individual electron-hole contributions. 
Exemplifying this, Fig.\ref{fig:qfactormain}(b) reports a comparison of the plasmonic peaks in the presence of three \textit{SW}, \textit{sv}, and \textit{dv} defects and their corresponding Lorentzian fits. After the fitting procedure, the quality factor (Q-factor), defined as the localized plasmon resonant energy over the FWHM, is computed. 
In particular, as reported in in Fig. \ref{fig:qfactormain}(c), we observe a decrease of the Q-factor upon insertion of defects in the pristine graphene nanostructure, whose data point corresponds to 0 number of defects (black dot), a behaviour that is maintained by increasing the number of defects until a kind of stationary value~\cite{Aguillon_2021} is achieved (around five or six defects). For better comparison, we have considered fixed positions for all the types of defects, as also the positions of the defects can affect the value of the Q-factor, the relative amplitude of individual data points [see Appendix (\ref{app:considerations})], as well as the steepness of the Q-factor decrease with an increasing number of defects. 
Furthermore, besides the variability linked to the position of the defects, we observe the following general trend: in the presence of \textit{sv}, the Q-factor tends to be higher than in the presence of \textit{dv} and \textit{SW}, and to decrease with a lower steepness. 
This is reported in Fig.\ref{fig:qfactormain}(c), showing the behaviour of the Q-factor with an increasing number of defects. The figure clearly highlights the significant difference in the Q-factor modification associated to the three kinds of defects. Specifically, the Q-factor calculated for the graphene nanostructure hosting 1 \textit{sv} can be as large as twice the Q-factor corresponding to 1 \textit{dv} or \textit{SW}. 
Similarly, the Q-factor associated to 2 \textit{SW} is equivalent to the Q-factor corresponding to 6 \textit{sv}.
\begin{figure}
\begin{overpic}[width=1.0\linewidth]{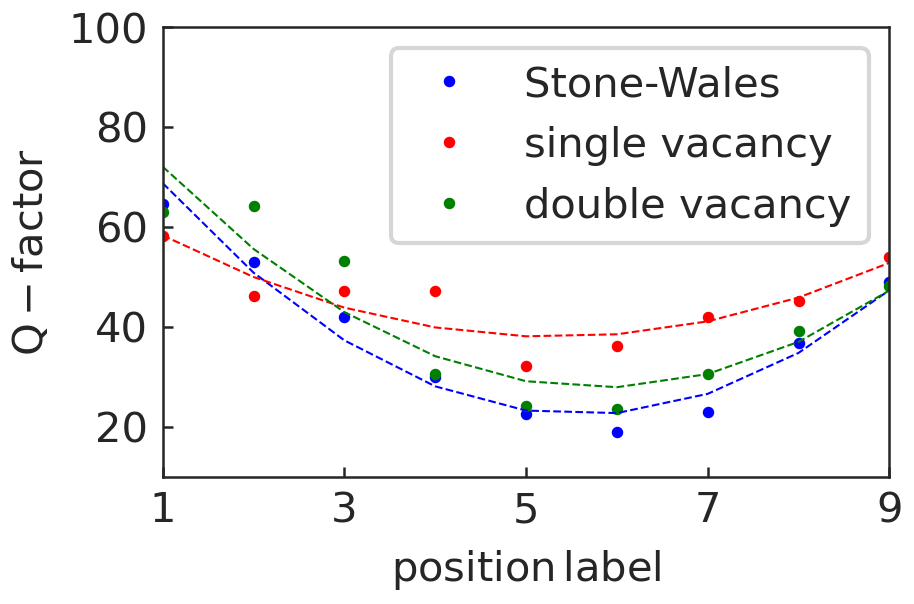}\put(2,67){(a)}\end{overpic}
\begin{overpic}[width=1.0\linewidth]{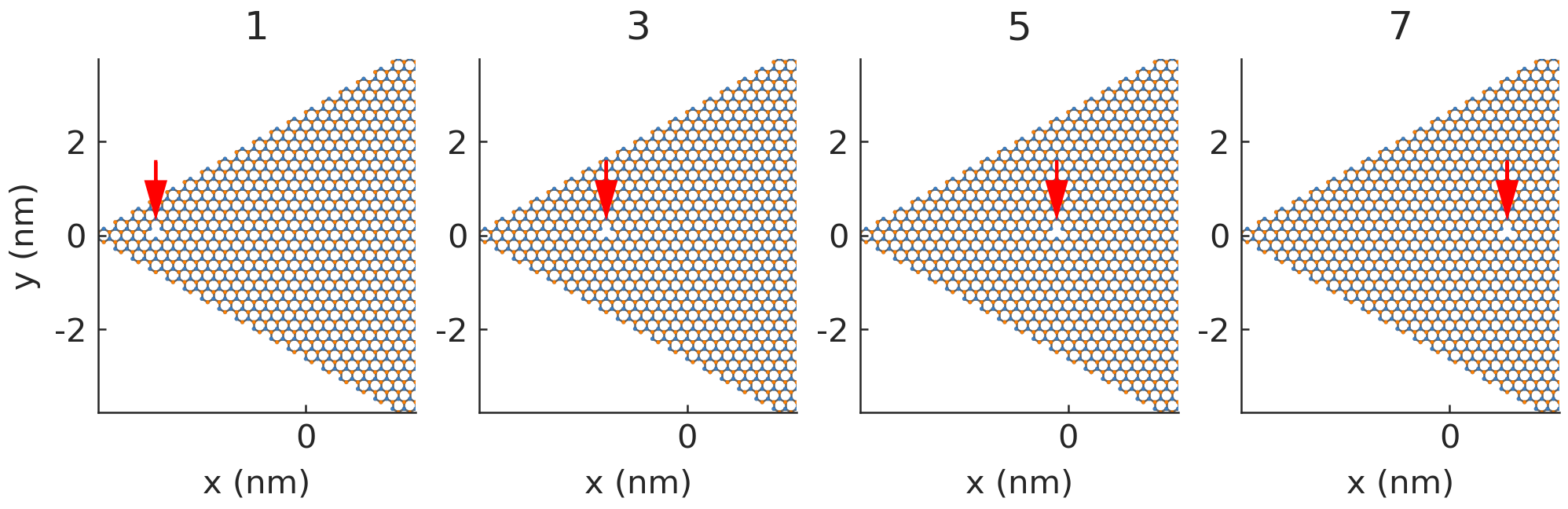}\put(2,32){(b)}\end{overpic}
\caption{\label{fig:qfactorchain}
Panel a) reports the quality factor of a single defect as a function of a position label which denotes defect positions $(x,y)$ ranging from the left to the right of the nanostructure along the $x$ direction, as shown in panel b). An increasing value of the position label indicates a higher value of the coordinate $x$, while $y$ is fixed. Panel b) atomic structure corresponding to labels $1,3,5,7,$ in panel (a). The red arrows highlight the defects. 
 }
\end{figure}
Finally, in Fig.~\ref{fig:qfactormain}, the position of the defects has been chosen randomly within the bulk of the nanostructure, in order to separate their effect from the effect associated to the edge-defects. 
This aspect is explored in Fig.~\ref{fig:qfactorchain}(a), reporting the behaviour of the Q-factor of a selected plasmonic peak of the graphene nanostructure decorated with one defect whose position ranges from edge to edge of the geometry, as shown in Fig.~\ref{fig:qfactorchain}(b). 
Interestingly, for all types of defects considered, a position within the bulk of the structure tends to correspond to a smaller Q-factor compared with the Q-factor related to a position that approaches an edge of the nanostructure. 
\begin{figure}[t!]
\begin{overpic}[width=1.0\linewidth]{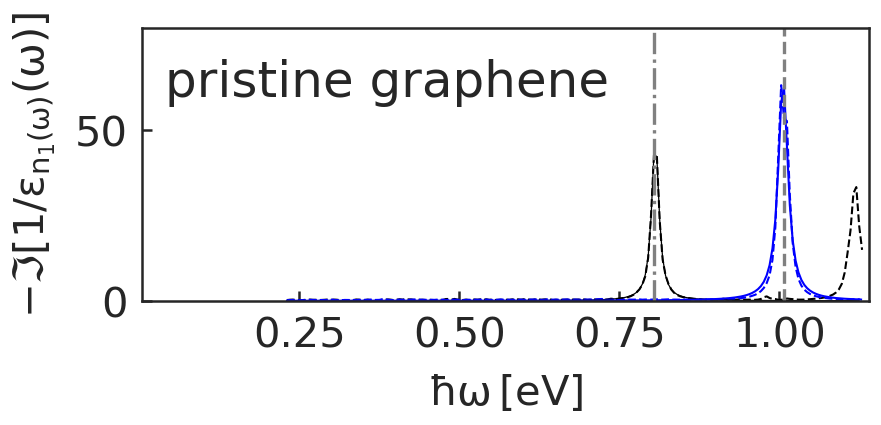}\put(2,51){(a)}\end{overpic}
\begin{overpic}[width=1.0\linewidth]{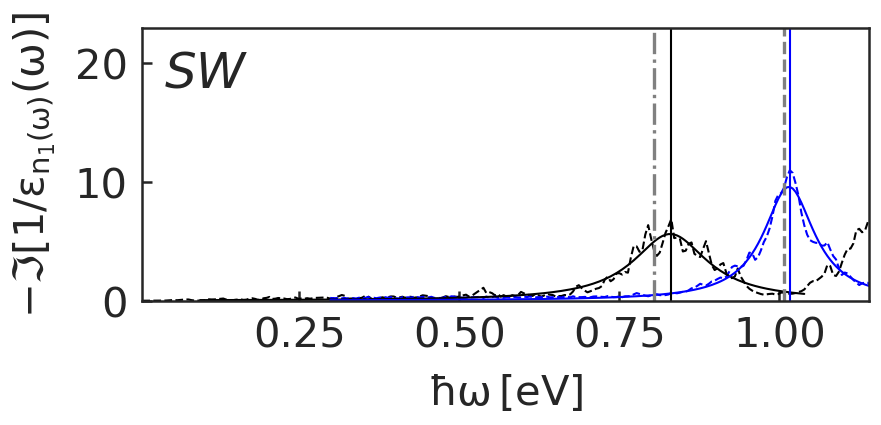}\put(2,51){(b)}\end{overpic}
\begin{overpic}[width=1.0\linewidth]{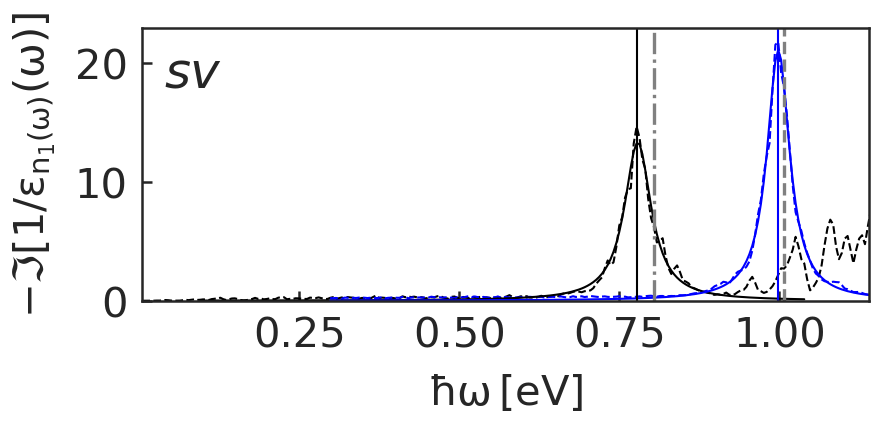}\put(2,51){(c)}\end{overpic}
\begin{overpic}[width=1.0\linewidth]{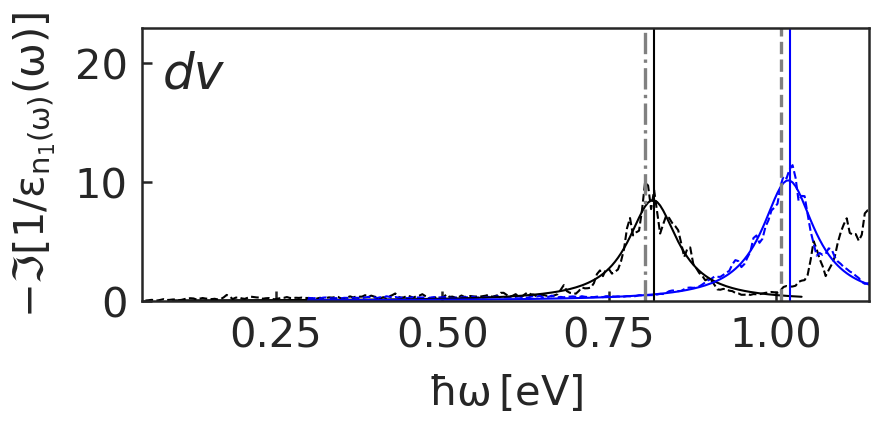}\put(2,51){(d)}\end{overpic}
\caption{\label{fig:spectratunability}
Spectra comparison of plasmon peaks calculated at $\mu = 1.0 \, {\rm eV}$ (black) and $\mu = 1.6 \, {\rm eV}$  (blue), corresponding to the pristine nanostructure (a) and to nanostructures hosting three defects of type \textit{SW} (b), \textit{sv} (c) and \textit{dv} (d). The dashed and dot-dashed vertical lines mark the energy of the plasmonic resonances of the pristine nanostructure. The solid vertical lines indicate the energy of the plasmonic resonances of the defective nanostructure. Both the plasmonic peaks (dashed curves) and the corresponding Lorentzian fits (solid curves) are reported. In (b), (c) and (d) the defects are located at the same positions and far from the nanostructure edges.}
\end{figure}

In the following, we qualitatively analyse the impact of \textit{SW}, \textit{sv} and \textit{dv} defects on the dynamic tunability of a targeted plasmonic resonance, quantified through the tuning range, defined as $T_{\rm R} = E_{\rm P}(\mu = 1.6 \, {\rm eV})-E_{\rm P}(\mu = 1.0 \, {\rm eV}) $, where $E_{P}(\mu)$ is the localized plasmon energy as a function of the chemical potential $\mu$.
This quantity was chosen as independent variable as the chemical potential, and therefore the localized plasmon energy, can be tuned by externally changing the gate voltage~\cite{Hu_2016}. 
In this respect and similarly to Ref.~\onlinecite{Jarzembski_2020}, in the calculation of $T_{\rm R}$ we have chosen a spectral window of $\approx 0.6 \, {\rm eV}$, even though we have set a higher chemical potential.
Indeed, a higher chemical potential provides greater resilience to a significant number of defects, thereby facilitating a meaningful comparison of plasmon resonances.
The energies $E_{P}(\mu)$, at different values of the chemical potential are computed from the Lorentzian fit of each plasmonic resonance, as reported in the example of Fig.~\ref{fig:spectratunability}, illustrating a comparison with the pristine case. For all the types of defects considered, at fixed defect number and type, $E_{P}(\mu)$ blue-shifts with increasing chemical potential, the same occurring in pristine graphene. At fixed chemical potential, $E_{P}(\mu)$ is found to red-shift with increasing \textit{sv} concentration, as it has been previously found in Ref.~\onlinecite{Aguillon_2021}, whereas the opposite trend is obtained for \textit{SW} and \textit{dv}. This observation hints that the different impact on the electronic structure induced by different kinds of defects affects directly the energy of the localized plasmon and its evolution with respect to defect concentration.
\begin{figure}
\begin{overpic}[width=1.0\linewidth]{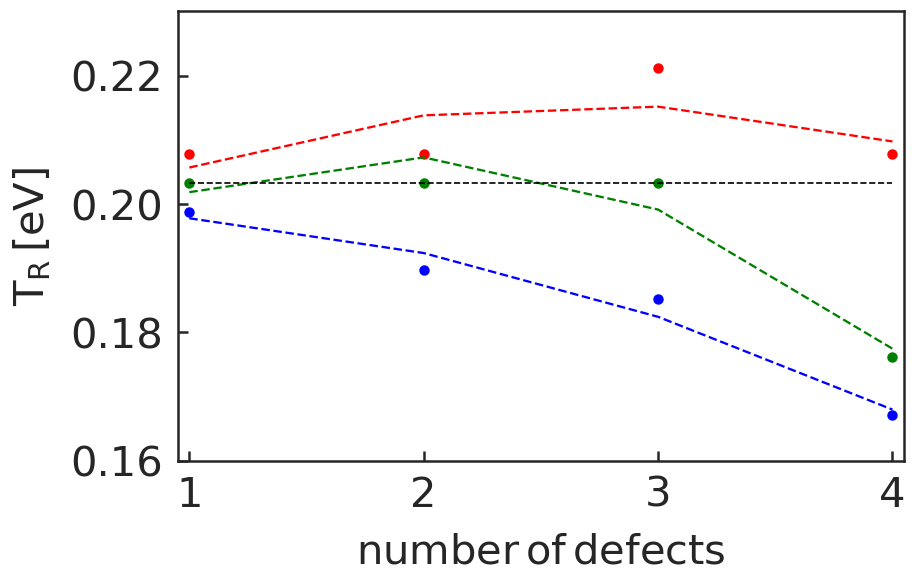}\end{overpic}
\caption{\label{fig:tunability}
Tuning range as a function of the number of defects. The \textit{SW} (blue), \textit{sv} (red) and \textit{dv} (green) defects are compared. The tuning range of the pristine nanostructure is displayed as a reference (black horizontal dashed line). The positions of the defects are the same for all three kinds of defects and are always far away from the nanostructure edges. The dashed lines are guides to the eye. 
 }
\end{figure}
In this respect, Fig.~\ref{fig:tunability} reports $T_{\rm R}$ in the presence of \textit{SW}, \textit{dv}, and \textit{sv} defects as a function of their number. When \textit{SW} defects are considered, the resulting tuning range is observed to be smaller than the corresponding value of the pristine nanostructure. Furthermore, the tuning range decreases with increasing the number of \textit{SW} defects. A similar behaviour is found to occur also for the \textit{dv} defects even though it is observed an appreciable change to the pristine tuning range only after 3 defects. In contrast, the nanostructure doped with \textit{sv} defects displays a non-monotonic tuning range. This is possibly related to the non-monotonic behaviour of the density of states at the Fermi level, hence of the plasmonic resonant energy, upon an increase of the number of the \textit{sv} defects~\cite{Aguillon_2021}.
In general, we observe that the presence of defects can induce an appreciable shift of $T_{\rm R}$ with respect to the pristine nanostructure. Here, the maximum shift induced by four \textit{SW} defects corresponds to $\approx 18 \%$ of the tuning range of the pristine nanostructure. 
Although preserving the possibility of setting the resonant energy of the localized plasmons by changing the chemical potential, defects influence the sensitivity to electrical control. Concretely, the magnitude of the gate voltage that needs to be externally applied to induce a desired shift of the localized plasmon energy may vary depending on the type of the hosted defect. For example, the presence of \textit{SW} is expected to limit the sensitivity to electrical control, that is, with respect to the pristine case, a given variation of the chemical potential induces a smaller variation of the plasmon energy.
In this respect and within the context of sensing schemes based on plasmonic enhancement, the variability of the tuning range due to the presence of defects is expected to show its major effects on i) SEIRA spectroscopy-based molecular sensors which rely on the dynamical tunability for molecular fingerprint identification~\cite{Hu_2016}, where the tuning of the plasmon energy is exploited to enhance the interaction with specific molecular vibrations; ii) molecular sensors that use the shift of the plasmon energy due to changes of the chemical potential, in turn induced by the presence of analyte molecules, directly for detecting and measuring the concentration of molecules~\cite{Bareza_2020}.
It is observed that besides the type of defect, also their number influences the behaviour of $T_{\rm R}$. The observed quantitative and qualitative changes of $T_{\rm R}$ hint that is important to take into account the presence of defects for evaluating the plasmonic response of graphene-based devices. Indeed, in order to fully utilize the benefits of the dynamical tunability, designing molecular sensors by taking into account the effect of defects---beyond established factors such as geometry, dielectric substrate, dielectric isolation of the substrate, that are already taken into account to optimize the dynamical tunability---could benefit the plasmon-sensing ability, in particular, improve the sensitivity of sensors based on graphene nanostructures.

\section{Concluding remarks}
Graphene-based nanostructures have emerged as highly promising platforms for developing molecular sensors that offer enhanced sensitivity, selectivity, and compactness. The plasmonic characteristics of pristine graphene serve as a fundamental advantage for molecular sensors in numerous proof-of-concept studies documented in existing literature. However, we demonstrate that the presence of structural defects in graphene nanostructures can significantly influence their plasmonic properties. In this study, we provide initial estimations of the quality factor and tuning range, which quantifies the dynamic tunability, and investigate their behavior in relation to the number and types of defects. Our calculations take into account previous findings in the literature, which have proposed methods for enhancing the plasmonic response of graphene-based nanostructures. Therefore, we consider several parameters to be crucial for our study: a high chemical potential, the minimum value for the relative permittivity of the dielectric environment, and a nanostructure geometry that exhibits greater resilience against carbon vacancies compared to other geometries. These choices of externally controllable operational parameters and device design components have proven advantageous for plasmon-based sensing, including SEIRA spectroscopy. While various types of defects are known to have distinctive effects on the electronic properties of graphene, their impact on plasmonic properties has not been adequately understood. Thus, we compare the influence of a low concentration of some of the most common structural defects in graphene, namely \textit{SW}, \textit{sv}, and \textit{dv}.
Our observations reveal that all analyzed defect topologies have a detrimental effect on the quality factor, which decreases as the number of defects increases. This behavior is governed by the broadening of the resonance and can vary depending on the specific type of defect. Despite the presence of defects, the ability to tune the plasmon energy by adjusting the chemical potential remains. However, defects alter the value of the pristine tuning range and the type of defect influences the relationship between the tuning range and the number of defects. The order of magnitude of the tuning range remains comparable to the pristine tuning range for up to four defects.
Considering that the quality factor directly correlates with the sensitivity of nanophotonic resonators, and the dynamic tunability indirectly determines the sensitivity by enhancing the matching with the targeted excitation of the analyte, and controls the operating energy range and spectral window of nanophotonic resonators, our findings can provide valuable insights for the development of molecular sensing based on graphene plasmonic enhancement in realistic environments. Specifically, our results emphasize the importance of considering changes in the plasmonic properties of a nanostructure induced by the presence of structural defects when designing a plasmonic platform for molecular sensing.
Furthermore, in light of recent demonstrations of the possibility of engineering the types of defects hosted by graphene~\cite{Lin_2020}, and given the extensive potential applications of graphene as a plasmonic material for molecular sensing, the investigation of the impact of structural defects on the plasmonic properties of graphene should be pursued experimentally to further enrich our understanding.

\acknowledgments
Part of this research has been funded by the Discovery Element of the European Space Agency.

\appendix
\section{FORMALISM} \label{app:formalism}
\begin{figure*}[t!] 
\begin{overpic}[width=0.49\linewidth]{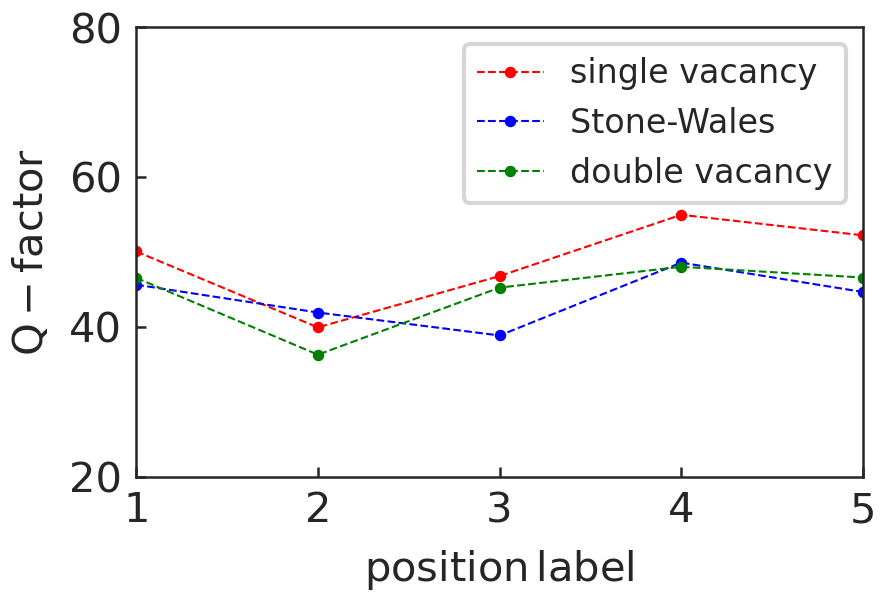}\put(2,67){(a)}\end{overpic}
\begin{overpic}[width=0.49\linewidth]{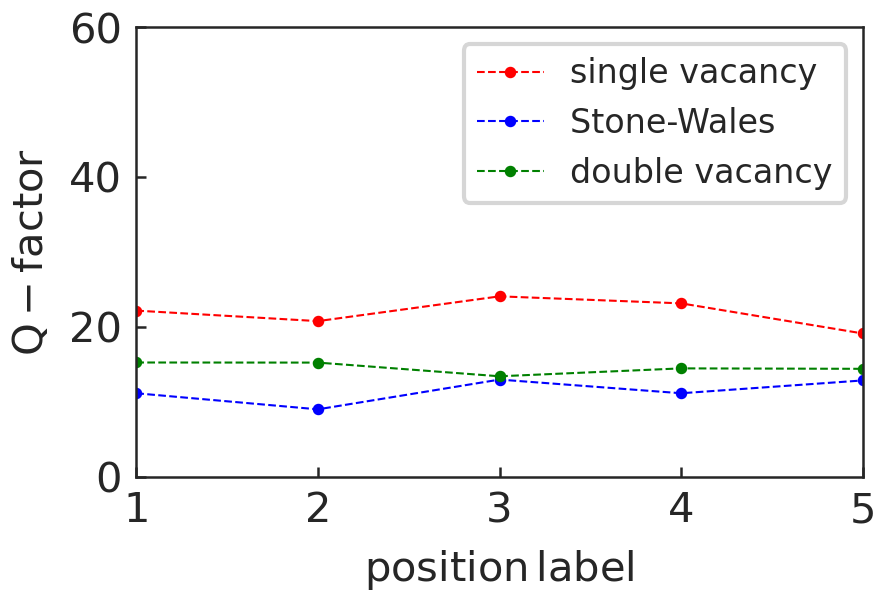}\put(2,67){(b)}\end{overpic}
\caption{\label{fig:variationwithpos}
Quality factor corresponding to randomly chosen positions denoted by a position label. Panel a) one defect, panel b) three defects. The range of the abscissa is the same in both panels for a better comparison. Dashed lines connect the data points.
 }
\end{figure*}
The method Q-TB+RPA consist on describing the dielectric function as a matrix in a basis of localized position eigenvectors $\langle  {\bf r} | $, 
\begin{equation} \label{eq:epsilon}
\langle {\bf r} | \hat{\varepsilon} (\omega) |{\bf r'} \rangle = \langle  {\bf r} | {\bf r' }\rangle -  \int d^{d} r^{''} \langle {\bf r}  | \hat{V_{\rm C}} |  {\bf r''} \rangle \langle {\bf r''} | \hat{\chi}(\omega)| {\bf r'} \rangle  
\end{equation}
in terms of the matrix elements of the Coulomb potential $\hat{V}_{\rm C} $, which is considered as a perturbation, and in terms of the matrix elements of the proper density-density response function in RPA approximation, namely, the non-interacting density-density response function $\hat{\chi}$. The matrix elements of the latter matrices are 
\begin{equation} \label{eq:coulomb}
\langle {\bf r} | \hat{V}_{\rm C} (\omega) |{\bf r''} \rangle \equiv \dfrac{e}{4 \pi \epsilon_{r}\epsilon_{0}|{\bf r}-{\bf r'' }|}
\end{equation}
and 
\begin{equation} \label{eq:chi}
\langle {\bf r''} | \hat{\chi}(\omega) |{\bf r'} \rangle = g_{s} \lim_{\eta \rightarrow 0+} \sum_{i,j}\langle i | \hat{G}|j \rangle  \langle j |{\bf r''}  \rangle \langle {\bf r''} |i \rangle \langle i |  {\bf r'} \rangle \langle {\bf r'}  | j \rangle
\end{equation}
respectively. The dimension of the calculation here is $d=2$. At vanishing distance $|{\bf r}-{\bf r'' }|$, the Coulomb interaction is considered to converge to the p-orbital onsite interaction energy $v_{0}=15.78 \, {\rm eV}$, in accordance with Ref.~\onlinecite{Thongrattanasiri_2012}. 
In equation (\ref{eq:chi}), $g_{s} = 2$ denotes the spin degeneracy, while the matrix with elements
\begin{equation} \label{eq:cgmatrix}
\langle i |\hat{G} |j \rangle \equiv \dfrac{n_{i}-n_{j}}{E_{i}-E_{j}-\hbar(\omega+ i \eta)}
\end{equation}
is defined in terms of the Fermi-Dirac distribution function,
\begin{equation} \label{eq:fddistri}
n_{i} \equiv \dfrac{1}{e^{(E_{i}-\mu)/k_{B}T}+1}
\end{equation}
determining the occupational number of the $i\rm{th}$ energy level, which depends on the eigenstates $\langle i |$ and on the eigenenergies $E_{i}$ of the tight-binding Hamiltonian $\hat{H}_{(0)} $.  
The tight-binding Hamiltonian specifies the model describing the electronic structure. Here we consider a minimal model, 
\begin{equation} \label{eq:tb-H}
\hat{H}_{(0)} = -\sum_{\langle lk \rangle} t_{lk} \hat{c}^{\dagger}_{l} \hat{c}_{k}
\end{equation}
where the summation is over nearest neighbors. From equation (\ref{eq:tb-H}),  the exact eigenstates and eigenenergies, inputs of the response function in equation  (\ref{eq:chi}), are obtained numerically. 
For the pristine graphene nanostructure, all carbon-carbon hopping energies $t_{lk}$ coincide with the bulk hopping energy  $t_{0} = 2.8 \, {\rm eV}$ and all carbon-carbon distances are $a_{\rm c-c} = 0.142  \, {\rm nm}$.
The parameter $t_{lk}$ is assumed to decrease exponentially on the carbon-carbon distance $r_{lk}$ after defect formation as 
\begin{equation} \label{eq:strainedhopping}
t(r_{lk})=t_{0}e^{- \beta [(r_{lk}/a_{\rm c-c})-1]}
\end{equation} \newline
where $\beta = 3.37$, in accordance with the strain model in Refs.~\onlinecite{Papaconstantopoulos_1998,Ribeiro_2009}.
From the numerical diagonalization of the dielectric function in equation (\ref{eq:epsilon}), 
we find the eigenvalues $\varepsilon_{n}(\omega)$ and the eigenvectors $|\phi_{n}(\omega)\rangle$ of the dielectric function, satisfying 
\begin{equation} \label{eq:epsilon_eigen}
\hat{\varepsilon} (\omega)|\phi_{n}(\omega)\rangle= \varepsilon_{n}(\omega)|\phi_{n}(\omega)\rangle \, .
\end{equation}
The diagonalization procedure is repeated for the range of frequency $\omega$ of interest.
Plasmon resonances are obtained from the loss function $-\Im[1/\varepsilon_{n_{1}}(\omega)]$, 
which is computed from the eigenvalues $\varepsilon_{n_{1}}(\omega)$ of the dielectric function that return the first maximum of such a functional of the eigenvalues, at each frequency $\omega$ of the spectrum considered. A peak in the loss spectrum is identified with a collective excitation (plasmon) when an eigenvalue $\varepsilon_{n_{1}}$ corresponding to a local maximum of the loss function satisfies $\Re[\varepsilon_{n_{1}}(\omega)]= 0$. 
Although we use the Q-TB+RPA approach to investigate graphene-based nanostructures, the method is neither restricted to this material nor to flat structures. The main limitation of the Q-TB+RPA method is its high computational cost, scaling as $N^{4}$, where $N$ is the number of atomic sites in the nanostructure. The computational cost has been recently lowered to $N^{3.13}$ in Ref.~\onlinecite{Westerhout_2022}.
\section{VARIABILITY OF THE Q-FACTOR WITH DEFECT POSITION} \label{app:considerations}
Fig.~\ref{fig:variationwithpos} reports an example of the Q-factor of the nanostructure considered in the main text, hosting one defect (a) and three defects (b), placed at different positions. The Q-factor is plotted against a position label that denotes a randomly chosen configuration of positions that excludes the edges of the nanostructure.
For all the three kinds of defects, the value of the Q-factor corresponding to the nanostructure hosting one defect appreciably changes with the position of the defect, Fig.~\ref{fig:variationwithpos}(a), while the variation of the Q-factor of the nanostructure hosting three defects tends to be less pronounced, Fig.~\ref{fig:variationwithpos}(b).   While the relative values of the Q-factor of the different kinds of defects might change for one defect, depending on their position, for the case of three defects the \textit{sv} defect returns the highest Q-factor, followed by the \textit{dv} and \textit{SW} defects. 

\end{document}